\documentclass[%
  reprint, 
  10pt,
  nofootinbib,
  amsmath,amssymb,
  aps,
superscriptaddress]{revtex4-2}

\usepackage{bm}
\usepackage{latexsym}
\usepackage{dcolumn}
\usepackage{amsmath,amsfonts,amssymb}
\usepackage{graphicx,epsfig}
\usepackage{psfrag}
\usepackage{amsthm}
\usepackage{color}
\usepackage{comment}

\usepackage{nccmath}
\usepackage{moresize}
\usepackage{enumerate}
\usepackage[hang,flushmargin]{footmisc}
\usepackage[titletoc,toc]{appendix}
\usepackage{lipsum}
\usepackage{caption}
\usepackage{subcaption}
\usepackage{epstopdf} 
\usepackage{hyperref}

\usepackage{pgfplots}
\usepgfplotslibrary{groupplots}
\pgfplotsset{compat=1.18}
\pgfmathdeclarefunction{asinh}{1}{\pgfmathparse{ln(#1 + sqrt((#1)*(#1) + 1))}}

\usepackage{rotating}
\usepackage{multirow}
\usepackage{mathtools}

\usepackage{stackengine}
\usepackage{scalerel}

\newcommand{\be}{\begin{equation}}
\newcommand{\ee}{\end{equation}}
\newcommand{\bea}{\begin{eqnarray}}
\newcommand{\eea}{\end{eqnarray}}
\newcommand{\bse}{\begin{subequations}}
\newcommand{\ese}{\end{subequations}}
\newcommand{\bce}{\begin{center}}
\newcommand{\ece}{\end{center}}
\newcommand{\bfg}{\begin{figure}}
\newcommand{\efg}{\end{figure}}
\newcommand{\bit}{\begin{itemize}}
\newcommand{\eit}{\end{itemize}}
\newcommand{\bed}{\begin{description}}
\newcommand{\eed}{\end{description}}
\newcommand{\ben}{\begin{enumerate}}
\newcommand{\een}{\end{enumerate}}
\newcommand{\nn}{\nonumber}

\newcommand{\fr}{\frac}
\newcommand{\sq}{\sqrt}

\def\a  {\alpha}

\def\D  {\Delta}

\def\l  {\lambda}
\def\L  {\Lambda}
\def\m  {\mu}
\def\n  {\nu}
\def\o  {\omega}
\def\O  {\Omega}

\def\th {\theta}

\def\s  {\sigma}

\def\vep {\varepsilon}

\newcommand{\cL}{\mathcal L}
\newcommand{\cM}{\mathcal M}

\newcommand{\cR}{\mathcal R}

\newcommand{\cV}{\mathcal V}
\newcommand{\cZ}{\mathcal Z}

\newcommand{\at}{\widetilde{a}}

\newcommand{\Geff}{G_{\text{\scriptsize eff}}}

\newcommand{\sw}{\mathsf w}

\newcommand{\bdm}{\begin{displaymath}}
\newcommand{\edm}{\end{displaymath}}

\begin{document}


\title{A Non-Singular Cyclic Universe from Modified Gravity}

\author{Saurya Das}
\email{saurya.das@uleth.ca}
\affiliation{
Theoretical Physics Group,
Department of Physics and Astronomy  \& Quantum Horizon Alberta, 
University of Lethbridge, 4401 University Drive,
Lethbridge, Alberta T1K 3M4, Canada 
}%
\author{Mitja Fridman}
\email{mitja.fridman@fjfi.cvut.cz}
\affiliation{
Department of Physics, Faculty of Nuclear Sciences and Physical Engineering, Czech Technical University in Prague, B\v{r}ehová 7, 115 19 Praha 1, Czech Republic 
}
\author{Dhiraj Kuniyal} 
\email{dhirajkuniyal@gmail.com}
\author{Sourav Sur}
\email{sourav@physics.du.ac.in}
\affiliation{
Department of Physics and Astrophysics, University of Delhi, Delhi 110007, India
}%


\begin{abstract}
The Big Bang singularity marks a fundamental breakdown of General Relativity, 
yet all proposed quantum gravitational 
remedies invoke microphysical assumptions beyond the observational reach. 
%
We show that a minimal, purely classical modification of the gravitational potential --- preserving the inverse-square law where tested, but free to depart at unmeasured scales --- is sufficient to produce a non-singular, cyclic universe. 
%
The resulting modified gravitational formulation leads to a cosmic evolution exhibiting a bounce at a finite minimum size, followed by a regime closely akin to the concordant $\Lambda$CDM model, and an eventual collapse to another bounce.
%
Cosmic Chronometer, Supernovae type Ia, Baryon Acoustic Oscillation, Cosmic Microwave Background and Nucleosynthesis
%
data constrain both the new length scales, implying a cycle period of 
%
${\simeq}\,89\,\mathrm{Gyr}$,
and this framework --- subsuming a broad class of bounce scenarios within a single testable model --- opens a direct route for distinguishing cyclic cosmologies from $\Lambda$CDM.
\end{abstract}
%
\maketitle


\section{Introduction} \label{sec:Intro}
Cosmological evolution in the standard realm is generally envisaged in several phases, as indicated by a wide range of observations, from primordial nucleosynthesis and the cosmic microwave background to the large-scale structure and the late-time accelerating expansion of the Universe. The base model, to which the observations broadly concord, is $\L$CDM, where $\L$ is a cosmological constant and CDM stands for cold dark matter.
Nevertheless, its backward extrapolation leads to the Big Bang singularity, where spacetime geodesics become incomplete and classical General Relativity (GR) ceases to be predictive. Resolving this singularity remains one of the central challenges of fundamental physics.

Many proposed resolutions invoke quantum gravitational effects, including string-inspired scenarios, loop quantum cosmology, and other non-singular bounce models. Although these approaches often replace the initial singularity with a finite minimum scale factor, they typically rely on specific microscopic assumptions whose observational consequences remain uncertain. It is therefore worthwhile to ask whether a non-singular cosmological evolution can emerge from a more general and phenomenological modification of gravity, independent of any particular quantum-gravity framework.

A useful starting point is the observation that gravity has been tested directly only over a finite range of length scales. Laboratory experiments constrain deviations from Newtonian gravity down to approximately 
%
a length 
$L_1 \sim 10^{-4}\,$m
\cite{submm1,kapner2007,adelberger2009,aspelmeyer}, 
while Solar-System, galactic, and cosmological observations probe scales extending up to at least  
a length 
$L_2 \sim 10^{26}\,$m
%
by taking account of the GR corrections to the Newtonian potential
%
\cite{bertottiCassini2003,will,planck2018}.
Outside 
%
the interval $(L_1,L_2)$ though,
%
the gravitational interaction remains unconstrained. 
%
This motivates us to look for a modification, or in some sense, 
%
a generalization of the Newtonian 
%
potential itself, preserving
%
all the 
%
experimentally tested predictions of Newtonian gravity and GR, however,
allowing smooth departures
%
from the latter
%
beyond the observationally accessible window.

%
\section{Modified Gravitational Potential} \label{sec:Mod_Grav}
\begin{figure}[t]
\centering
\includegraphics[width=0.98\columnwidth]{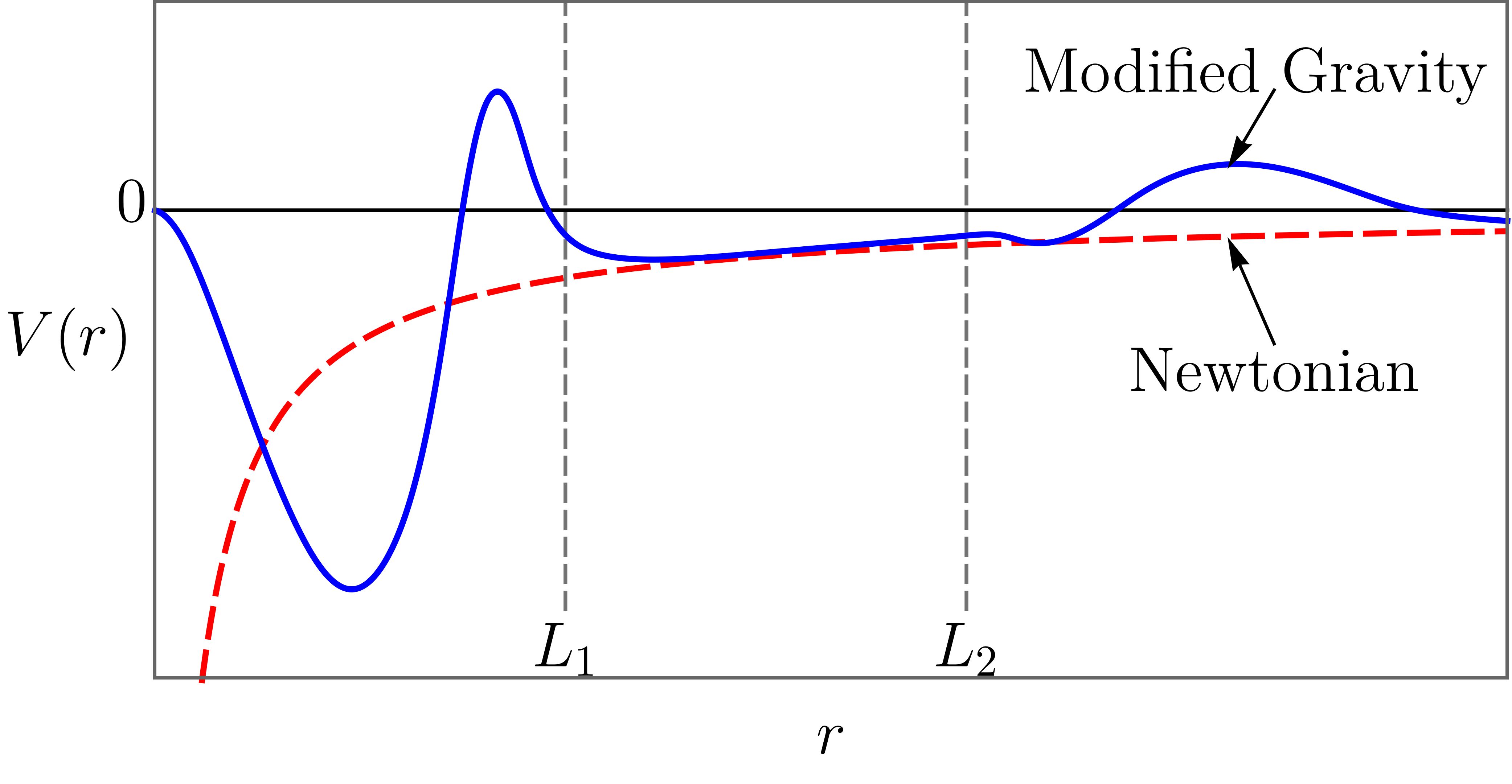}
\caption{
Schematic exemplary functional variation of the modified gravitational potential $V(r)$; with the range $r \in (L_1, L_2)$ being that constrained by observations.
}
\label{fig:vr6}
\end{figure}

%
Let us consider the following modified form of the Newtonian gravitational potential for a point mass $M$:
\be \label{cond0}
V(r) = - \fr{GM} r \, F(r) \,, 
\ee 
where $F(r) = 1$ for $r \in (L_1, L_2)$, but is in general a function of $r$ that parametrizes possible departures from the realm of standard gravity beyond $(L_1, L_2)$, as depicted by the illustrative schematic in Fig.\,\ref{fig:vr6}. By `standard gravity' we mean Newtonian gravity as well as GR, the weak-field limit of which recovers the Newtonian potential.
The same weak-field limit recovery can in principle be demanded for Eq.\,(\ref{cond0}) as well, which can thus be embedded in a relativistic (covariant) theory of gravity.  
Nevertheless, without a prior specification of such an embedding, 
a suitable choice of $F(r)$ can lead to a broad class of non-singular cosmologies, including the bouncing and cyclic ones, which we shall demonstrate in due course. In fact, this forms our mainstream work in this paper, alongside the examination of the observational viability of the emerging scenario.
%
%
%
%
%

%
Now, a generic way of parametrizing departures from the standard gravitational realm 
is to factorize
\be \label{F}
F(r) =  F_1(r) \cdot F_2(r) \cdot F_3(r) \,,
\ee
where $F_1(r)$, $F_2(r)$ and $F_3(r)$ are three functions which can be chosen suitably to play specific roles as follows:
%
%

The function $F_1(r)$ can be chosen such that the singularity at $r=0$ gets {\em smoothened}. A suitable choice is
\be \label{F1}
F_1(r) = e^{- \ell_1/r} \fr{\left(r/\ell_2\right)^\a}{1 + \left(r/\ell_2\right)^\a} \,,
\ee 
where $\a$ is a dimensionless positive constant, and $\ell_1, \ell_2$ are positive constants
of the dimension of length, either of which could be the Planck length, however, strictly speaking, 
%
%
they just need to be less than $L_1$.

%
The function $F_2(r)$, on the other hand, can make the gravitational potential $V(r)$ deviate further from the Newtonian one, outside the experimentally tested range $r \in (L_1, L_2)$, without the introduction of any zeroes however. A suitable choice for this may be, for instance, 
\be \label{F2}
F_2(r) = \left[1 + \sum_{l=1}^s \left(\frac{r_{0l}} r\right)^l\right] \left[1 + \sum_{l'=1}^{s'}\left(\frac r {r'_{0l'}}\right)^{l'}\right] \,,
\ee 
where $r_{0l}$ and $r'_{0l'}$ are two sets of constants of dimension of length, $r_{0l} < L_1$ and $r'_{0l'} > L_2$ for all integer values of the indices $l$ and $l'$ respectively, up to certain finite numbers, $s$ and $s'$. 
%
%
Assuming, without loss of generality,
$r_{0s} > r_{0 (s-1)} > \dots > r_{01}$
and
$r'_{0s'} > r'_{0 (s'-1)}> \dots > r'_{01}$,
one can distinguish the nature of $V(r)$ at
%
%
%
%
%
short and long distances, respectively. 

%
The function $F_3(r)$ can finally be chosen to incorporate zero(es) in $V(r)$. A suitable chosen form is 
\be \label{F3}
F_3 (r) = \prod_{i=1}^n \left(1 - \fr{r_i}r\right)^{\nu_i} \prod_{j=1}^{n'} \left(1 - \fr r {r_j'}\right)^{\nu_j'} ,
\ee 
which encodes all the short-distance zeroes $r_i \ll L_1$ (say, $n$ in number) and all the long-distance zeroes $r'_j\gg L_2$ (say, $n'$ in number), 
with respective integer multiplicities $\nu_i$ and $\nu'_j$, under the assumption
$r_n < r_{n-1} < \dots < r_1$ and 
$r'_1 < r'_2 < \dots < r'_{n'}$,
without loss of generality.
%
%
%

In the remainder of 
%
this paper, we shall stipulate the constants 
$\ell_1 = \ell_2 = 0$, as well as assume $r_{0l} \to 0$ and $r'_{0l'} \to \infty$, 
%
whence $F_1(r) = F_2(r) = 1$, i.e., the full function $F(r) = F_3(r)$ given by Eq.\,(\ref{F3}).
Moreover, we shall set $n = n' = \nu_1 = \nu_1' = 1$ in Eq.\,(\ref{F3}), since as explained below in section\,\ref{sec:Cyc}, a single pair of short and long distance {\em simple} zeroes (i.e., the zeroes of unit multiplicity) is necessary and sufficient for a non-singular cyclic universe.  
%

\section{Cosmological Setup} \label{sec:Cosm}
Given the modified Newtonian potential $V(r)$, Eq.\,(\ref{cond0}), as our starting point, it is convenient to resort to a quasi-Newtonian (QN) cosmological formulation, following standard texts
\cite{inverno,weinberg,liddle}. 
In particular, we take the specific approach of ref.
\cite{DS-VarG},
applicable to the standard form of the gravitational potential except with a varying Newtonian coupling $\Geff(r)$. This is tenable since we can always recast Eq.\,(\ref{cond0}) in such a form with $\Geff(r) = G F(r)$.

With $F(r) = F_3(r)$, 
alongside $n = n' = \nu_1 = \nu_1' = 1$ in Eq.\,(\ref{F3}), the QN approach yields the following dynamical equation for a system of $N$ perfect fluids (presumably non-interacting):
%
\be \label{MFE1}
H^2 
= \fr{8 \pi G}{3 c^2} \sum_{I=1}^N \fr{\rho_{_{I0}}}{a^{3 (1 + \sw_{_I})}} 
\left(1 - \fr{a_1} a\right) \left(1 - \fr a {a_1'}\right) - \fr{k c^2}{a^2} \,,
\ee 
where 
$\,a = a(t) = r(t)/\ell_0$ (with $\ell_0$ a fiducial length) is the cosmological scale factor, $H = H(t) = \dot a(t)/a(t)$ (with the dot $\{\cdot\} \equiv d/dt$) is the QN analogue of the Hubble parameter, $a_1$ and $a_1'$ are the small and large values of $a$ at which $H$ vanishes, i.e., the early {\em bounce} and future {\em turnaround} points in the course of evolution of the Universe respectively,
$\rho_{_{I0}}$ are the energy densities of the individual fluid species at the present epoch ($t = t_0$), $\sw_{_I}$ are the corresponding equation-of-state (EoS) parameters (considered to be constants, for brevity), and $k$ is a dimensionless constant proportional to the total energy of the system per unit mass, $E$ (see the Appendix \ref{App:QN_Cosm} for all the details). Thus, $k = 0$ when $E = 0$, and furthermore, one can always set $k = \pm 1$ for $E \lessgtr 1$, irrespective of stipulating the scale factor $a = 1$ at the present epoch ($t = t_0$) without loss of generality. Hence, from a purely mathematical standpoint, Eq.\,(\ref{MFE1}) can be interpreted as the QN analogue of the Friedmann equation in standard cosmology, with the last term in the former resembling the spatial curvature term appearing in the latter. 
%
%
%
%

In fact, Eq.\,(\ref{MFE1}) may rather appropriately be referred to as the modified Friedmann equation (MFE), arising from the modified Newtonian potential $V(r)$. 
Note also that the QN approach, although quite heuristic, generically leads to cosmological evolution equations same in mathematical form as those describing the cosmic background level dynamics in the standard framework
\cite{DS-VarG}.
This is irrespective of the component fluid/field configuration, even when many such components are relativistic {\it per se}. 
%
Therefore, notwithstanding the QN origin, 
Eq.\,(\ref{MFE1}) can safely be taken as the governing equation in a cosmological analysis concerning the background evolution. 
On the other hand, the analysis that inherently refers to the cosmological perturbations would require a relativistic (covariant) embedding of the modified Newtonian potential. 
%
This, despite being a hard proposition, is shown to be achievable 
in some of our works --- the preceding ones
\cite{DS-VarG,DFS-SFG},
and that currently underway
\cite{SKD-Cov}
--- see the discussions in the concluding part of this paper.
%

For simplicity, we consider only the known forms of fluid components, viz., radiation, non-relativistic pressureless matter (dust) and a cosmological constant $\L$, with the respective energy densities 
$\rho_{r0}$, $\rho_{m0}$ and $\rho_{\L0}$
at the present epoch ($t = t_0$ or $a = 1$), and EoS parameters
$\sw_r = 1/3$, $\sw_m = 0$ and $\sw_\L = -1$.
Defining, in the usual way, the corresponding density parameters, as well as that for the $k$-contributor, at $t = t_0$:
\bea \label{dens_par}
&& \O_{r0} = \fr{\rho_{r0}}{\rho_{c0}} \,, \quad
\O_{m0} = \fr{\rho_{m0}}{\rho_{c0}} \,, \quad
\O_{\L0} = \fr{\rho_{\L0}}{\rho_{c0}} = \fr{\L c^2}{3 H_0^2} \,, \nn \\
&& \O_{k0} = - \fr{k c^2}{H_0^2} \,,
\eea 
with $\rho_{c0} = 3 c^2 H_0^2/(8 \pi G)$ the critical density at $t = t_0$, and $H_0 \equiv H \vert_{t = t_0}$ the Hubble's constant, we express Eq.\,(\ref{MFE1}) rather conveniently as
\bea \label{MFE2}
H^2 &=& H_0^2 \left[\left(\fr{\O_{r0}}{a^4} + \fr{\O_{m0}}{a^3} + \O_{\L0}\right) \right. \nn\\
&& \times \left(1 - \fr{a_1} a\right) \left(1 - \fr a {a_1'}\right) + \left.\fr{\O_{k0}}{a^2}\right] .
\eea
Evaluating this at $t = t_0$ (or $a = 1$) we get the constraint
\be \label{constr}
\O_{r0} + \O_{m0} + \O_{\L0} = \fr{1 - \O_{k0}}{\left(1 - a_1\right) \left(1 - a_1'^{-1}\right)} \,,
\ee
using which we eliminate $\O_{\L0}$ from Eq.\,(\ref{MFE2}) to write 
\bea \label{MFE3}
\fr{H^2}{H_0^2} &=& \fr{\O_{k0}}{a^2} + \bigg[\O_{r0} \left(\fr 1 {a^4} - 1\right) + \O_{m0} \left(\fr 1 {a^3} - 1\right)   \nn\\
&+& \fr{1 - \O_{k0}}{\left(1 - a_1\right) \left(1 - a_1'^{-1}\right)}\bigg] \left(1 - \fr{a_1} a\right) \left(1 - \fr a {a_1'}\right).~~
\eea
Check that for $a_1 = 0$ and $a_1' \to \infty$, Eqs.\,(\ref{MFE2}),\,(\ref{constr}) and (\ref{MFE3}) reduce to the respective ones for the $\L$CDM model in presence of radiation and the $k$-term. However, given the minuscule abundance of the Cosmic Microwave Background (CMB) radiation in the observed Universe at present, we shall ignore the contribution of $\O_{r0}$ in what follows. Furthermore, we shall set $\O_{k0} = 0$, in view of the gross observational support for the spatial flatness in the standard cosmological paradigm. 

\section{Parameter Estimates and comparison with $\L$CDM} \label{sec:Param}
With $\O_{r0} = 0 = \O_{k0}$, Eq.\,(\ref{MFE3}) reduces to the following form, in terms of the redshift $z = a^{-1} - 1\,$:
\bea \label{MFE4}
\fr{H^2(z)}{H_0^2} &=& \bigg[\O_{m0} \left\{(1 + z)^3 - 1\right\} + \fr 1 {\left(1 - a_1\right) \left(1 - a_1'^{-1}\right)}\bigg] \nn\\
&& \times \Big[1 - a_1 (1 + z)\Big] \Big[1 - a_1'^{-1} (1 + z)^{-1}\Big] .
\eea
In order to confront this with observational data, we carry out a statistical maximum likelihood analysis for the parameters concerned, using the standard Markov-chain Monte Carlo (MCMC) technique.
%

The datasets we consider are from three low-$z$ observations and two high-$z$ ones. The low-$z$ data include $51$ direct measurements of $H(z)$ using (primarily) the Cosmic Chronometers (CC)
\cite{Moresco2016,Monjo2024},
$1701$ datapoints from the Pantheon$^+$ Supernovae type-Ia (SNIa) distance-redshift measurements 
\cite{Brout2022}, 
and $15$ datapoints from the baryon acoustic oscillation (BAO) distance measurements 
\cite{Eisen2005,eBOSS2020,DESI2024}. 
The high-$z$ data, on the other hand, comprise of the compressed Planck 2018 CMB distance-prior likelihood ($4$ points)
\cite{Arendse2020}, 
and the primordial light-element abundance constraints ($2$ points) from big bang nucleosynthesis (BBN) 
\cite{Pitrou2021,Aver2015,Cooke2018}
--- see the Appendix \ref{App:Lik_Method} for more details. As explained therein, a joint analysis with all these data amounts to minimizing the total negative log-likelihood, or $\chi^2_{\rm tot}$, with respect to the parameters concerned.

The parameter vector for our cyclic scenario is
%
\be \label{par-vec}
\boldsymbol{\th} = \left\{H_0, \O_{m0}, \D M, r_d, \o_b, N_1, N_1'\right\} ,
\ee 
where, apart from $H_0$ and $\O_{m0}$, we have the nuisance parameter $\D M$ denoting the absolute-magnitude offset of the Pantheon$^+$ SNIa sample, the comoving sound horizon at the baryon drag epoch, $r_d$, the physical baryon density at the present epoch, $\o_b$,
and the bounce and turnaround scale factor ten-folds, $N_1 \equiv \log_{10} a_1$ and $N_1' \equiv \log_{10} a_1'$, respectively. 
For $\L$CDM, which we consider as a reference model to compare with, all the parameters except $N_1$ and $N_1'$ in Eq.\,(\ref{par-vec}) are of relevance. The total number of parameters, $N_{\rm par}$, is therefore, five for $\L$CDM, compared to the seven in the cyclic scenario.
%
\begin{figure*}[htb]
\centering
\includegraphics[width=0.78\textwidth]{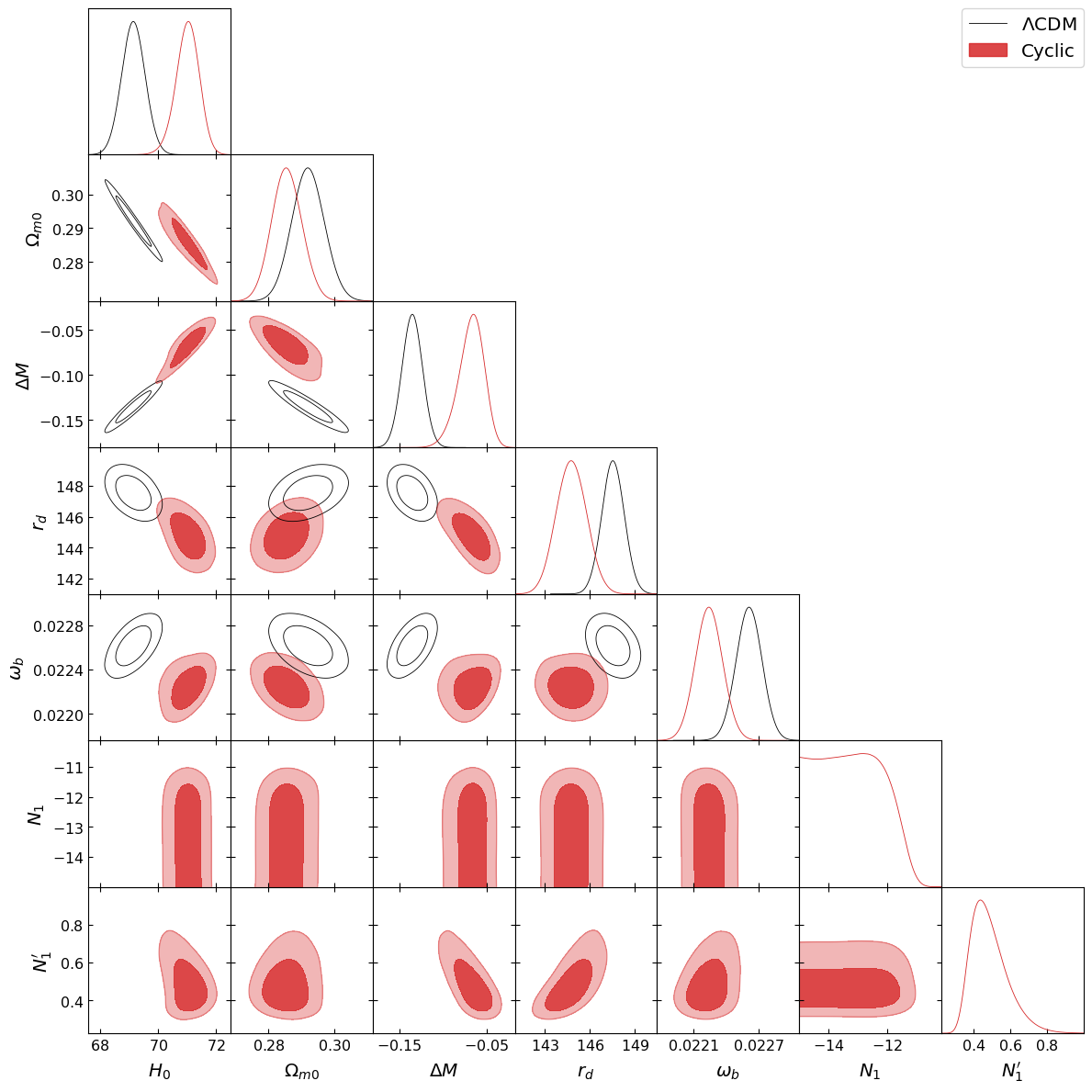}
\caption{Marginalized one-dimensional and two-dimensional ($1\s$ and $2\s$) posteriors for the cyclic scenario vis-\`a-vis $\L$CDM from the joint analysis using the CC\,+\,Pantheon$^+$\,+\,BAO\,+\,CMB\,+\,BBN data. 
}
\label{fig:1D2D_post}
\end{figure*}
%
\begin{table}[t]
\centering
\renewcommand{\arraystretch}{1.5}
%
\caption{Priors and estimates of parameters (upto the $68\%$ confidence level) for the cyclic scenario and $\L$CDM.
} 
\begin{tabular}{lccc}
\hline\hline
Parameter & Prior & Cyclic & $\L$CDM \\
\hline
$H_0 ({\rm km/s/Mpc})$ & $[50,90]$
& $71.0267_{-0.4031}^{+0.3821}$
& $69.1491_{-0.3997}^{+0.4026}$ \\

$\O_{m0}$ & $[0,0.5]$
& $0.2855_{-0.0045}^{+0.0046}$
& $0.2921_{-0.0049}^{+0.0050}$ \\

$\D M$ & $[-1,1]$
& $-0.0670_{-0.0154}^{+0.0133}$
& $-0.1351_{-0.0117}^{+0.0116}$ \\

$r_d~({\rm Mpc})$ & $[120,180]$
& $144.7630_{-0.9923}^{+1.0090}$
& $147.5370_{-0.7384}^{+0.7463}$ \\

$\o_b$ & $[0.018,0.026]$
& $0.0222_{-0.0001}^{+0.0001}$
& $0.0226_{-0.0001}^{+0.0001}$ \\

$N_1$ & $[-15,-9]$
& $-13.2296_{-1.2030}^{+1.1910}$
& --- \\

$N_1'$ & $[0,1]$
& $0.4703_{-0.0797}^{+0.1122}$
& --- \\

\hline \hline
\end{tabular}
\label{tab:constr}
\end{table}
%

Fig.\,\ref{fig:1D2D_post} shows the one- and two-dimensional posterior distributions (up to $2\s$) for the cyclic scenario as well as for $\L$CDM, obtained using the Goodman-Weare affine-invariant ensemble sampler {\it emcee} for MCMC
\cite{GdWr2010},
with flat priors listed in Table\,\ref{tab:constr}, alongside the estimated parameter values (best fits and the $1\s$ or $68\%$ confidence limits). The convergence is assessed by the well-known Gelman-Rubin statistic
\cite{GlRb1992},
with the associated tolerance factor found to be ${\widehat R} = 1.011$ for the cyclic scenario and ${\widehat R} = 1.0025$ for $\L$CDM. While the former is well within the range $(1.01,1.05)$ categorized `good', the latter is `excellent' ($< 1.01$).
%

%
\begin{table*}[htb]
\centering
\renewcommand{\arraystretch}{1.7}
\caption{
Comparative statistics, including the goodness-of-fit, information criteria, and the Bayesian evidence, for the cyclic scenario and $\L$CDM. The changes, denoted by $\D$, refer to the differences from the $\L$CDM values.
}
\begin{tabular}{lccccccccc}
\hline\hline
Model & $N_{\rm par}$ & $\chi^2_{\min}$ & $\chi^2_{\rm dof}$ & AIC & $\D$AIC & BIC & $\D$BIC &
$\ln \cZ$ & $\D \ln \cZ$ \\
\hline
Cyclic & $7$ & $1828.29$ & $1.0353$ & $1842.29$ & $-27.29$ & $1880.65$ & $-16.33$ & $-938.34$ & $13.30$ \\
$\L$CDM & $5$ & $1859.58$ & $1.0518$ & $1869.58$ & $0.00$ & $1896.98$ & $0.00$ & $-951.65$ & $0.00$ \\
\hline\hline
\end{tabular}
\label{tab:comparison}
\end{table*}
%
%
%
%

The estimated best fit $N_1 = - 13.2296$ and the best fit $N_1' = 0.4703$ in Table\,\ref{tab:constr}
correspond to the 
scale factor values $a_1 = 5.8939 \times 10^{-14}$ and $a_1' = 2.9532$ respectively. While the later suggests a future turnaround from an expanding to a contracting phase of cosmic evolution when the Universe grows to about thrice as much its present size, the former shows that the bounce in the past, from a contracting to the expanding phase, occurs when the Universe is extremely minute in size. In fact, such a low value of $a_1$ implies that the bounce happens
well before the epoch at which cosmological perturbations associated with the observed CMB anisotropies enter the horizon, 
%
or in other words,
%
all the observable CMB modes remain super-horizon during the bounce. 
%
Hence, we may safely infer that the cosmic background evolution due to the MFE keep the CMB power spectrum unaffected 
and the cyclic scenario 
consistent with the Planck CMB observations. 
%
Also, the bounce is far upstream of the primordial nucleosynthesis which typically occurs around $a \sim 10^{-10}$. So the MFE modification to the standard form of the early expansion history of the Universe during the BBN era, till the recombination, is practically negligible. As such, the standard pre-recombination or BBN microphysics, including the sound horizon and the physics encoded in the compressed Planck likelihood remain unchanged, thus ensuring the compliance of the cyclic scenario with the BBN constraints as well.
%
%

%
Table\,\ref{tab:constr} also quotes a higher best fit value of the Hubble constant $H_0$ for the cyclic scenario than for $\L$CDM, $71.0267$ to $69.1491$. This may be considered as a positive outcome since in a joint analysis of any given model, using data from both low-$z$ and high-$z$ observations, an $H_0$ estimate higher than its $\L$CDM value is often desired for a potential resolution of the Hubble tension (see for e.g.
\cite{DiValentino2021}
and references therein). While no claim for the complete aversion of the discrepancy amounting to such a tension can be made as such for the cyclic scenario, we may say that the latter at least opens a phenomenological avenue for alleviating the persistent problem. In fact, this is corroborated by another desirable outcome --- the lowered best fit of $r_d$, the sound horizon at the photon-drag epoch, for the cyclic scenario than for $\L$CDM, $144.763$ to $147.537$ in Table\,\ref{tab:constr}. 

On the whole, in spite of having two additional parameters, the cyclic scenario provides a better fit to the combined data than $\L$CDM. This is shown by the corresponding statistics summarized in Table\,\ref{tab:comparison}, viz., the minimized $\chi^2_{\rm tot}$ which is denoted by $\chi^2_{\rm min}$, the goodness of fit $\chi^2_{\rm dof}$ which is the $\chi^2_{\rm min}$ per degree of freedom (dof), the Akaike and Bayesian information criteria, AIC and BIC, and the Bayes factor $\ln \cZ$ (see the Appendix\,\ref{App:Lik_Method} for  definitions). There are significant reductions in the $\chi^2_{\rm min}$, $\chi^2_{\rm dof}$, AIC and BIC values, and a substantial enhancement in $\ln \cZ$, for the cyclic scenario than those for $\L$CDM --- all of which imply better fitting of the former with the data compared to the latter. In particular, the changes $\D$AIC $= - 27.29$ and $\D$BIC $= -16.33$ are significant enough to favor the cyclic scenario over $\L$CDM comprehensively. A fairly large Bayes factor $\D \ln \cZ = 13.3$, computed via nested sampling
\cite{Skilling2006},
supports this further, and in fact shows a strong evidence in favor of the cyclic scenario on the Jeffreys' scale
\cite{Jeffreys1961}.
\section{Cyclic evolution and the $\L$CDM tracking phase}
\label{sec:Cyc}


Recall the MFE\,(\ref{MFE1}) above, or more specifically Eq.\,(\ref{MFE2}),
describing the cyclic evolution 
for the minimal setting $n = n' = \n_1 = \n_1' = 1$ in Eq.\,(\ref{F3}). Such a setting is in fact quite distinct, rather than a simplifying one. To see this, note first that Eq.\,(\ref{F3}), and consequently the cosmological equation\,(\ref{MFE0}) in the Appendix\,\ref{App:QN_Cosm}, imply the following generalized form of Eq.\,(\ref{MFE2}):
\be \label{MFEgen}
\fr{H^2}{H_0^2} = \left(\fr{\O_{m0}}{a^3} + \O_{\L0}\right) \prod_{i=1}^n \!\left(1 - \fr{a_i} a\right)^{\nu_i} \prod_{j=1}^{n'} \!\left(1 - \fr a {a_j'}\right)^{\nu_j'} ,
\ee
assuming $\O_{r0} = 0 = \O_{k0}$. Consider next, a particular bounce point $a = a_b$, which is a zero of multiplicity $\n$, say. In a close neighborhood (say, $\vep$) of $a_b$, one has $\dot a \propto (a - a_b)^{\n/2}$. So the traversal time from $a_b$ to $a_b + \vep$, i.e.,  
$ 
\D t \propto \int_{a_b}^{a_b + \vep} da \left(a - a_b\right)^{-\n/2} \,,
$
converges only for $\n = 1$. Similar argument holds in the vicinity of a particular turnaround $a = a_t$ as well. Thus, {\em simple} zeroes, i.e., the zeroes of multiplicity $\n = 1$, are {\em necessary}, and
for any $\n \geq 2$ the bounce (turnaround) degenerates into an asymptotically static state approached over infinite cosmic time --- the emergent (loitering) universe limit --- and the cyclicity is lost.
%
Moreover, since $H^2 > 0$ throughout, in the interval between a given bounce $a_b$ and a given turnaround $a_t$, the cosmic evolution oscillates indefinitely, thus forming multiple cycles. Nevertheless, our predictions on the past or(and) future are based on what we see in our present observable universe, at $t = t_0$ or $a = 1$. Therefore, limiting our attention to one zero on each side of $a = 1$, pertaining to a single cycle, is {\em sufficient}. All other cycles are mere replica of this cycle, which we label for convenience as `$1$'. This means that the corresponding bounce-turnaround pair  
%
$(a_1,a_1')$, together with the measured density parameters, 
determines the 
%
evolving scale factor, $a(t)$, at {\em all} times.
%

In the integral form Eq.\,(\ref{MFE2}) is expressed as
\bea \label{MFEint}
t - t_0 &=& \fr 1 {H_0} \int_1^a \fr{d\at} \at \left[\left(\fr{\O_{r0}}{\at^4} + \fr{\O_{m0}}{\at^3} + \O_{\L0}\right) \right. \nn\\
&& \times \left(1 - \fr{a_1} \at\right) \left(1 - \fr \at {a_1'}\right) + \left.\fr{\O_{k0}}{\at^2}\right]^{-1/2} .
\eea 
With $\O_{r0} = 0 = \O_{k0}$, and
$\O_{\L0}$ eliminated via the constraint\,(\ref{constr}), we work out the integral numerically by using the estimated values of $\O_{m0}$, $a_1$ and $a_1'$, to obtain $H_0 (t - t_0)$ as a function of the scale factor $a$ in the cyclic scenario. Similar functional variation is found for $\L$CDM, by setting $a_1 = 0, a_1' \to \infty$ and using the corresponding estimate of $\O_{m0}$. Inverting these we determine the time-evolution of the scale factor, $a(t)$, and consequently that of the Hubble parameter, $H(t)$, in either cases. 

%
\begin{figure}[htb]
\centering
\includegraphics[width=0.975\linewidth]{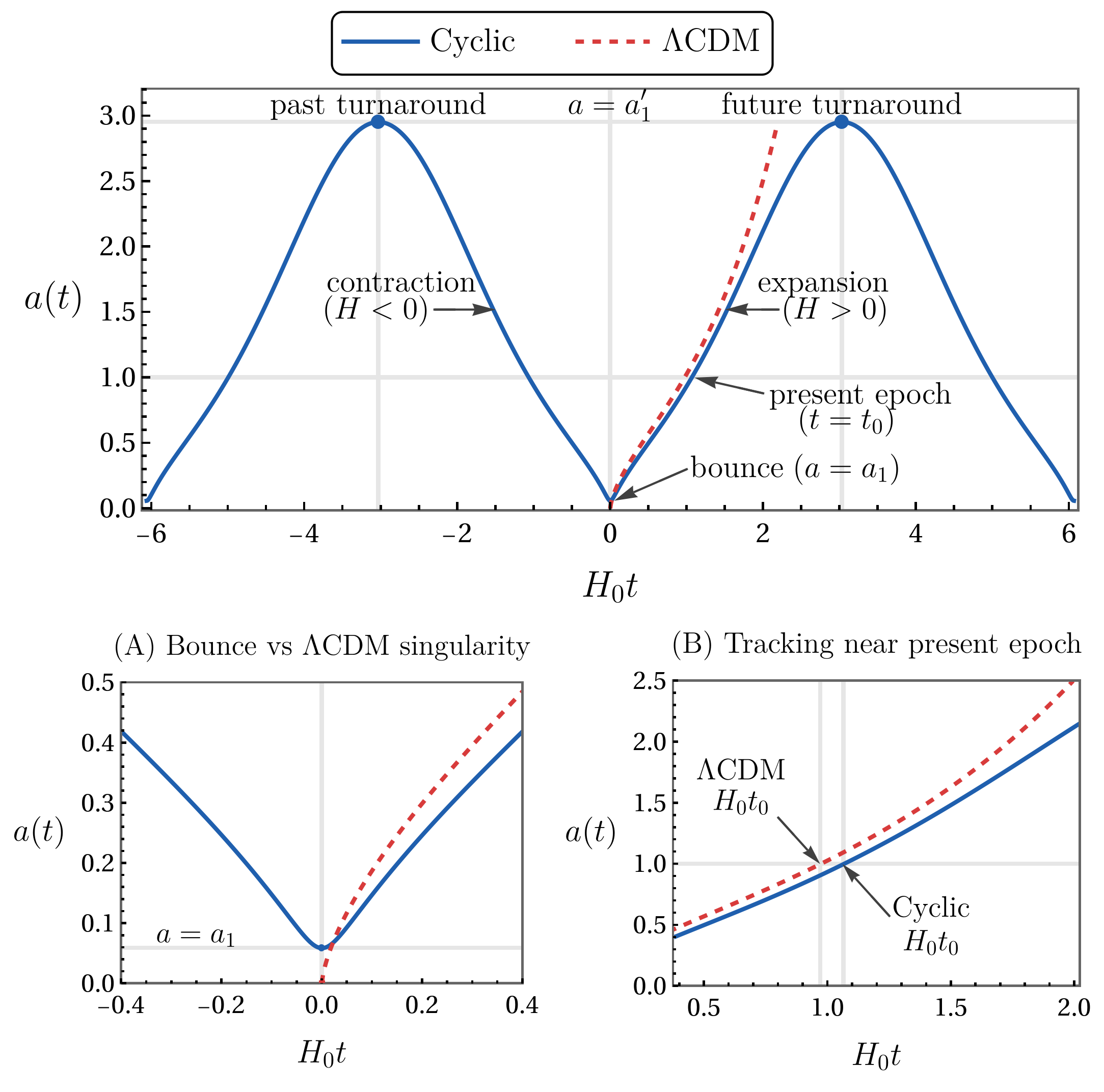}
\caption{[Top] Time-evolution of the cosmic scale factor, $a(t)$, in the cyclic scenario and in the $\L$CDM model. \newline [Bottom] Exaggerated depiction near (A) the cyclic bounce at $a = a_1$ and the $\L$CDM singularity at $a = 0$, and (B) the present epoch ($a = 1$). 
For illustration, the estimated $a_1$ is enhanced by a factor of $10^{12}$.}
\label{fig:evolving_a(t)}
\end{figure}
\begin{figure}[htb]
\centering
\includegraphics[width=0.95\linewidth]{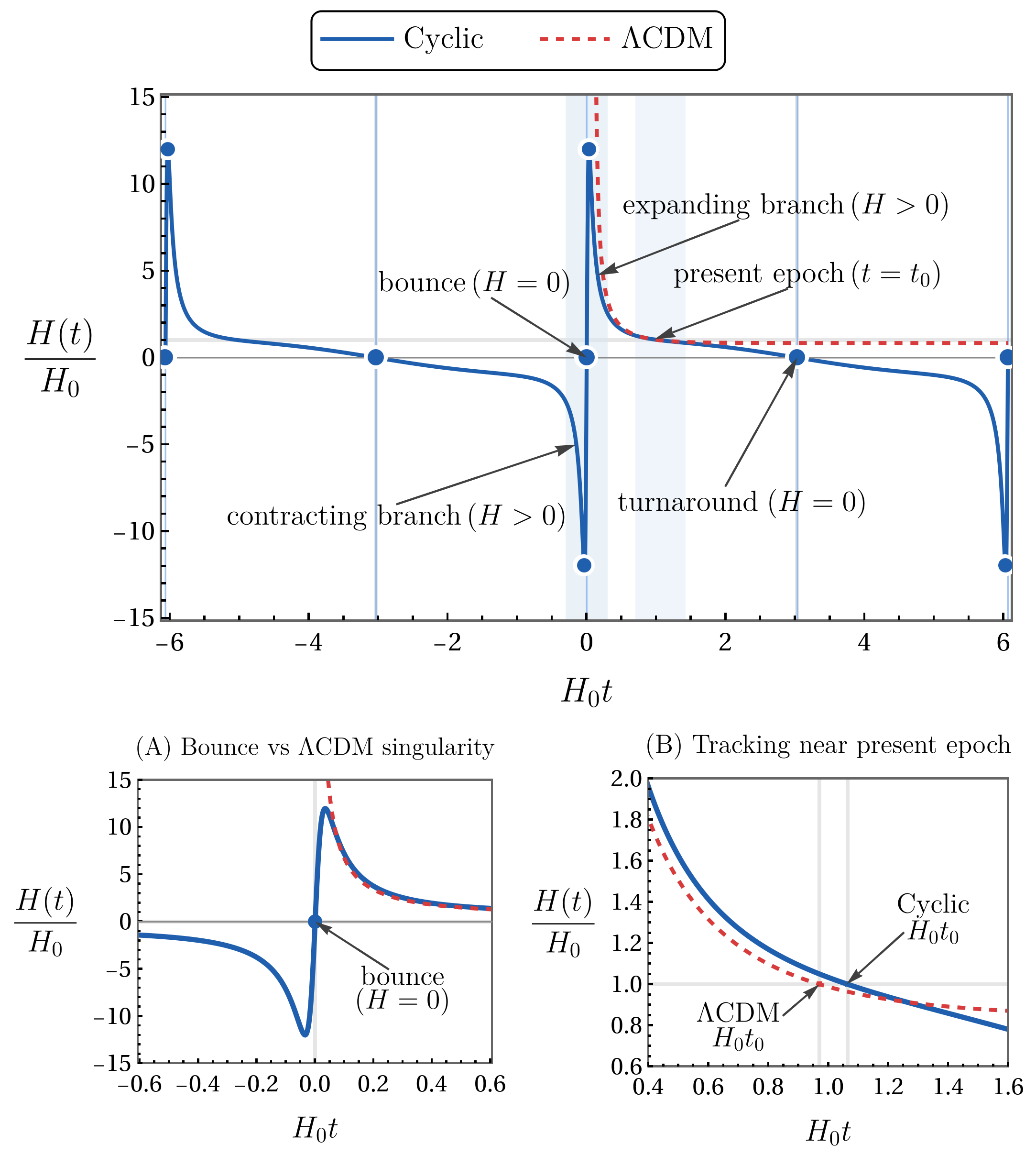}
\caption{[Top] Time-evolution of the Hubble parameter, $H(t)$, in the cyclic scenario and in the $\L$CDM model. \newline [Bottom] Exaggerated depiction near (A) the cyclic bounce at $a = a_1$ and the $\L$CDM singularity at $a = 0$, and (B) the present epoch ($a = 1$). 
For illustration, the estimated $a_1$ is enhanced by a factor of $10^{12}$. }
\label{fig:evolving_H(t)}
 \end{figure}
%

While the evolution of $a(t)$ with the dimensionless time variable $H_0 t$ (or $t/t_H$, where $t_H = H_0^{-1}$ is the Hubble time) is shown in the top panel of Fig.\,\ref{fig:evolving_a(t)}, that of the reduced Hubble parameter $H(t)/H_0$ with $H_0 t$ is shown in the top panel of Fig.\,\ref{fig:evolving_H(t)}. In each of these figures, the evolution before and after the bounce 
are depicted by the cycles on the left and right respectively, and the contracting and expanding phases, the present epoch and the bounce and turnaround points are duly marked. In fact, the  value of the scale factor at the bounce, $a = a_1$ (taken to be its estimated best fit, $5.8939 \times 10^{-14}$ in Table\,\ref{tab:constr}), is enhanced by a factor of $10^{12}$ in order to make the illustration clear. Moreover, exaggerations near $a = a_1$ and near $a = 1$ are shown, respectively, in the bottom panels (A) and (B) of Fig.\,\ref{fig:evolving_a(t)} and of Fig.\,\ref{fig:evolving_H(t)} for further clarity.

Right after the bounce, the cyclic evolution begins to track the $\L$CDM one, but slowly drifts away as time progresses. The drift  remains quite insignificant however, implying that the tracking continues even at the present epoch $t = t_0$, and afterwards till a cosmic time $t \simeq 14\, H_0^{-1}$ or so. Of course, the value of $t_0$, and hence the present age of the Universe, differs in the cyclic and $\L$CDM cases. While for $\L$CDM, the estimated best fit values of the associated parameters listed in Table\,\ref{tab:constr} show that the present age of the Universe (since the big bang) is 
$
0.9713\,H_0^{-1}|_{\L{\rm CDM}} = 13.7440\,{\rm Gyr} \,,
$
for the cyclic Universe, the associated best fit parametric estimates show that the present age (since the bounce) is 
$
1.0651\,H_0^{-1}|_{\rm Cycl} = 14.6727\,{\rm Gyr} \,.
$
The same estimates also indicate that the next turnaround is at $29.8165\,{\rm Gyr}$ from the present epoch, and the full cosmic cycle has a period of $88.9784\,{\rm Gyr}$.

%

\section{Generality of the scenario} \label{sec:Gen_Scn}

It is worth emphasizing that Eqs.\,(\ref{MFE1})--(\ref{MFE2}) 
%
cannot simply be regarded as those of another bouncing cosmological model expanding the literature on the subject. They actually describe
a {\em class} of cosmologies  
%
governed by a generalized form of the Friedmann equation given by
%
\be \label{Gen_FE}
H^2(a) = \fr{8\pi G}{3c^2}\,\rho(a)\,g(a)\,,
\ee
where $\rho(a)$ is the energy density of the total fluid content of the Universe, and the function
%
%
%
$g(a)$ collects whatever microphysical or gravitational corrections the underlying theory
%
provides.
The total fluid content may in general be due to several components. For $N$ such components with energy densities $\rho_{_{I0}}$ at the present epoch, and (presumably constant) EoS parameters $\sw_{_I}$, the expression for $\rho(a)$ is that given by Eq.\,(\ref{fluid_density}) in the Appendix\,\ref{App:QN_Cosm}. 
%

A non-singular bounce at $a=a_b$
%
and a subsequent turnaround at $a = a_t$ occur where $H=0$ and $\dot H\,>\,0$ and $<\,0$ respectively,
%
i.e., where $g(a)$ has positive real zeroes at a
%
small value of $a$ and at a large value of $a$, respectively. A natural form of $g(a)$ in terms of
%
its short-distance roots $a_i$ (corresponding to $r_i \ll L_1$) and large-distance roots $a_j'$ (corresponding to $r_j' \gg L_2$) 
%
is the factorized structure of Eq.\,(\ref{MFEgen}), i.e., 
\be \label{g}
g(a) = \prod_{i=1}^n \left(1 - \fr{a_i} a\right)^{\nu_i} \prod_{j=1}^{n'} \left(1 - \fr a {a_j'}\right)^{\nu_j'} \,,
\ee
with $\nu_i$ and $\nu_j'$ denoting the multiplicities of the zeroes.
%
Specific bounce and cyclic scenarios differ only in 
%
the characterization of
%
{\em which} $g(a)$ their dynamics generate.
Therefore, all such scenarios live inside the family parametrized by 
%
$(\{\rho_{_{I0}}\},\{\sw_{_I}\},\{a_i\},\{a_j'\},\{\nu_i\},\{\nu_j'\})$
%
%
\cite{brandenbergerBounceReview}.
Within this family, finite-time cyclicity selects the simple zeroes uniquely, and the innermost zeroes 
%
at each end of a given cycle
%
controls the entire accessible evolution. The minimal choice $n = n' = \nu_1 = \nu_1' = 1$ is therefore not a simplifying assumption but the observationally sufficient representative of the whole class.

Let us cite a few examples:

\medskip
\noindent
1. {\it Loop quantum cosmology} 
%
(LQC)\,: The formulation 
effectively leads to a Friedmann equation of the form
\be \label{MFELQC}
H^2 = \fr{8 \pi G}{3 c^2} \,\rho \left(1 - \fr{\rho}{\rho_c}\right) \,,
\ee 
%
%
with $\rho_c$ of the order of the Planck density
\cite{ashtekarSinghLQC,ashtekarPawlowskiSinghPRL,taverasLQC}. 
%
%
Assuming the Universe to be composed of a solitary perfect fluid with a constant EoS parameter $\sw$, we have $\rho = \rho_0 \, a^{- 3 (1 + \sw)}$, where $\rho_0$ is a constant. A comparison of Eqs.\,(\ref{Gen_FE}) and (\ref{MFELQC}) then shows
\be \label{LQC_g}
g(a) = 1 - \left(\fr{a_b} a\right)^{3 (1 + \sw)} , ~ \text{with} ~~~ a_b = \left(\fr{\rho_0}{\rho_c}\right)^{1/[3 (1 + \sw)]} ,
\ee 
implying the realization of 
%
%
%
a single short-distance {\em simple} zero of 
$H(a)$, i.e., a bounce, in LQC. No other zero exists, so there is 
%
no turnaround.

\medskip
\noindent
2. {\it Braneworld cosmology} :\;
%
Scenarios emerging from the braneworld models of the Randall--Sundrum type generically modify the Friedmann equation to
\be \label{MFEBC}
H^2 = \fr{8 \pi G}{3 c^2} \,\rho \left(1 \pm \fr{\rho}{\rho_c}\right) \,,
\ee 
with the $\pm$ in the brackets set by the sign of 
%
the brane tension $\l$,
%
and $\rho_c \sim 2|\l|$ 
%
\cite{shtanovSahniBrane,randallSundrum2,binetruyBraneFriedmann}. 
%
For a negative $\l$, i.e., with the minus sign in the brackets, we have the factor $g(a)$ same as that given by Eq.\,(\ref{LQC_g}), 
and hence a short-distance simple zero (bounce).
%
For a positive $\l$, however, the factor $g(a) > 0$, i.e., with  
%
no real root and the evolution remains singular.
%
On the whole, the
%
zero-structure is thus identical to that in LQC, 
%
although it originates from a higher dimensional geometry, via the application of the brane junction conditions, rather than from a four dimensional modified (quantum) gravitational potential. 
%

\medskip
\noindent
%
3. {\it $f(R)$ cosmology} :\; Formulations based on metric $f(R)$ theories of modified gravity, where $R$ denotes the Ricci curvature scalar, lead to a Friedmann equation precisely of the form given by Eq.\,(\ref{Gen_FE}), with the function $g(a)$ encoding
%
the curvature back-reaction, 
%
i.e., the effect of the terms in $f(R)$ that modify or correct the standard Einstein-Hilbert Lagrangian, $R$, on the space-time structure.
While explicit forms involving short distance (or ultra-violet) corrections, such as $f(R) = R + \a R^2$, where $\a$ is a dimensionful constant,  
%
reproduce the LQC-type bounce factor (\ref{LQC_g}) in the high-curvature limit 
%
\cite{nojiriOdintsovFR,deFeliceTsujikawa},
%
those with infrared corrections, for e.g., $f(R) = R - \m^4/R$, where $\m$ is another dimensionful constant, can lead to a zero of $H(a)$ at a large value of $a$, i.e., a future turnaround
%
%
\cite{carrollFR}. 
Thus $f(R)$ gravity naturally hosts both the elements of the 
zero-structure
%
of cyclic cosmology
%
within a single classical framework.

\medskip
\noindent
%
%
%
4. {\it Quintom matter-bounce cosmology} :\; 
Zero(es) of $H(a)$ can also arise from the cancellation of the total (positive) fluid density, $\rho(a)$, 
by an additional cosmic component with an effective negative energy density, $- \rho_{_X}(a)$.
Such a component can most simply be envisaged as due to the minimal gravitational coupling to a phantom-like quintessence scalar field or a {\em quintom}.
Several matter-bounce models have effective quintom constructions 
%
%
\cite{caiBounceReview,caiQuintomBounce,caiSaridakisQuintomReview}, 
%
resulting in a Friedmann equation of the form given by Eq.\,(\ref{Gen_FE}), with $g(a) = 1 - \rho_{_X}(a)/\rho(a)$. Assuming once again, for simplicity, $\rho = \rho_0 \, a^{- 3 (1 + \sw)}$, and also a constant EoS parameter $\sw_{_X}$ for the quintom, so that $\rho_{_X} = \rho_{_{X0}} \, a^{- 3 (1 + \sw_{_X})}$, we have
\be \label{Quint_g}
g(a) = 1 - \left(\fr{a_\ast} a\right)^{3 (\sw_{_X} - \sw)} , ~ \text{with} ~~~ a_\ast = \left(\fr{\rho_{_{X0}}}{\rho_0}\right)^{1/[3 (\sw_{_X} - \sw)]} ,
\ee 
where $\rho_{_{X0}}$ is a constant. Therefore, depending on the values of $\sw_{_X}$ and $\sw$, a {\em simple} zero can be perceived at a small $a$ (implying a bounce) or at a large $a$ (implying a turnaround). In fact, there can be many such simple zeroes in rather complicated situations --- that of a multi-component $\rho$ or(and) a multi-component $\rho_{_X}$. 
%

\medskip
\noindent
%
5. {\it Ekpyrotic cosmology and the cyclic extension} :\;
A very long slow contracting phase of cosmic evolution, before a bounce to an expanding one, can be driven effectively by a scalar field  
%
with a steep {\em negative} potential, 
%
as in the original {\em ekpyrotic} model 
%
\cite{khouryEkpyrotic,steinhardtTurokCyclic,lehnersEkpyrotic}. 
%
While the contraction due to the negative potential corresponds to an EoS parameter $\sw \gg 1$ in the equivalent fluid description, the bounce is postulated to be resulting from a collision of two four dimensional branes in a higher dimensional space, on one of which our physical world resides. The branes may be in an eternal oscillatory motion implying that the collision, and hence the bounce, may repeat itself periodically, thus forming cycles in the cosmic evolution. Now, at
the background 
%
cosmological
%
level, a genuinely cyclic 
%
evolution
%
requires {\em two} accessible simple zeroes of 
%
the function $g(a)$ in Eq.\,(\ref{Gen_FE}) --- at a small $a$ and at a large $a$ (the bounce and turnaround points respectively). This is precisely the structure we show in this work,
%
with repeated cycles given by the periodic extension of $a(t)$. 
The ``new ekpyrotic'' (ghost-condensate) variant,
%
which
%
achieves the bounce via null-energy-condition violation rather than a potential barrier, 
%
indeed
%
maps to the same short-distance simple zero in $g(a)$
\cite{buchbinderKhouryOvrut,arkaniHamedGhost}.

\medskip
Across these examples, 
%
one can see that
%
the model-dependent physics enters 
%
in Eq.\,(\ref{Gen_FE}) only through the specification of the
%
location, number, and order of the zeroes of 
%
the modulating function $g(a)$. 
With $g(a)$ given by Eq.\,(\ref{g}), it is evident that Eq.\,(\ref{Gen_FE}) is characteristically equivalent to the Fridmann equation describing
%
a unified, model-independent parametrization of 
%
not just the bouncing, but cyclic cosmologies as well, where
%
$\{a_i\}$ and $\{a_j'\}$ 
%
are
%
the phenomenological parameters to be confronted with 
%
the observational 
%
data.

\section{Conclusion} \label{sec:Concl}
Let us conclude with a discussion on the following:

\medskip
\noindent
%
1. {\it Covariant formulation}: 

\vspace{3pt}
\noindent
%
The basis of our analysis in this paper is the MFE given by Eq.\,(\ref{MFE1}), 
which we have derived in a quasi-Newtonian (QN) setup, rather than the standard cosmological one. Admittedly, we preferred a direct application of the modified Newtonian potential $V(r)$ given by
Eq.\,(\ref{cond0}), over the longer procedure of embedding $V(r)$ first in a relativistic (covariant) theory of gravity and then carrying out the cosmological analysis.
The purpose is served since in general the QN evolution equations are of the same mathematical form as in standard (relativisitic) cosmology, but at the background level only. At the level of the density perturbations,
%
the QN formalism
differs considerably from the relativistic one, and is in some sense untenable. 
%
So there is always a need for an equivalent covariant formulation. In fact, the covariant embedding of the potential $V(r)$ is in fact essential for a 
cosmological analysis concerning CMB, BAO, etc., which have direct bearing on the density perturbations.
%
Such an embedding can be achieved from diverse perspectives, as in our preceding works
\cite{DS-VarG,DFS-SFG}.
However, not all of them would
allow $V(r)$ to vanish, i.e., the function $F(r)$ in Eq.\,(\ref{cond0}), and hence the effective Newtonian coupling $\Geff(r) = G F(r)$ to vanish, anywhere. For instance, in ref.
\cite{DS-VarG}
a direct replication of the QN cosmological equations, for a given form of $V(r)$, is argued from the local limit of the non-local teleparallel equivalent of General Relativity 
\cite{BCHM-NLG,TBM-NLG},
via a suitable choice of the gravitational {\em susceptibility} function therein. Such a function is linear in $\Geff^{-1}$ and finite, meaning that $\Geff$ is strictly non-zero. 
Nevertheless, there is a fairly straightforward, yet unrestrictive way --- which is to make the legitimate demand that 
%
any covariant gravitational theory whose static, spherically symmetric, weak-field limit reproduces $V(r)$  
would yield
%
the Friedmann equation\,(\ref{Gen_FE}) in general,
%
at the leading order in a cosmological setting, with sub-leading 
corrections entering at higher order in $H^2/M_P^2$,
%
where $M_P = \sq{c/(8 \pi G)}$ is the reduced Planck mass.
%
The covariant embedding thus effected in the $f(R)$ theory, as an example, is illustrated in ref.
\cite{DFS-SFG},
for $V(r)$ of a specified functional form (different from the one here though). 
The generic embedding of $V(r)$, of any given form, in $f(r)$ gravity forms a major part of an ongoing work
\cite{SKD-Cov},
in an endeavor to a complete covariant formulation along this line. In as much, with the above stipulations for $V(r)$ considered here, the $f(R)$ action can lead to the same $g(a)$ zero-structure of Eq.\,(\ref{g}).  
%
Crucially, the observational constraints place the bounce and turnaround well away from the Planck scale --- $a_1 \simeq 5.89 \times 10^{-14}$ and $a_1' \simeq 2.95$ --- so the sub-leading covariant corrections are suppressed and do not alter this zero-structure on which the cyclic evolution rests. 
%
%
A full analysis of cosmological perturbations in this context, particularly the study of the
%
propagation of modes through the bounce and the pre-bounce initial-conditions, 
%
is a projected future work 
\cite{DFKS-Pert},
consequent to 
\cite{SKD-Cov}.
%

\medskip
\noindent
%
2. {\it Aversion of the entropy problem}:

\vspace{3pt}
\noindent
A classic objection to cyclic cosmologies, dating back to Tolman
\cite{tolman1934}, 
is that the entropy generated within each cycle accumulates, driving successive cycles to larger amplitude and longer period, so that a truly periodic history is impossible.
%
The cyclic scenario we have studied here
evades the amplitude part of this argument, 
%
as
%
the bounce and turnaround scales $a_1$ and $a_1'$ are fixed by the zeroes of the gravitational potential, not by the matter content, 
%
and therefore the
%
amplitude cannot grow from cycle to cycle. Entropy production can 
%
%
however
%
lengthen the period by changing the effective 
%
%
abundance of the cosmic constituents 
%
between the cycles. 
%
The requisite
thermodynamic treatment --- including possible dilution of entropy during the accelerated phase preceding the turnaround, as invoked in other cyclic scenarios
\cite{steinhardtTurokCyclic}
--- 
%
nevertheless requires the completion of the covariant formulation currently underway
\cite{SKD-Cov},
and is hence deferred to a future work.
%
%

\bigskip
\begin{acknowledgments}
This work was supported by the Natural Sciences and Engineering Research Council of Canada.
SS acknowledges financial support from the Faculty Research Programme Grant -- IoE, University of Delhi (Ref.No./IoE/2025-26/12/FRP).
\end{acknowledgments}

\medskip
\noindent\textbf{Data availability.}\;
There are no new observational data  generated for this work --- all the data used are taken from the cited references and available in the associated public data release domains.
%
%
%
%
%
%
%
%
%
%


\medskip
\noindent\textbf{Author contributions.}\;
All authors contributed equally to this work, including the conception of the project, the
theoretical framework, the analytic and numerical calculations, the cosmological data analysis,
and the writing of the manuscript.

\medskip
\noindent\textbf{Competing interests.}\;
The authors declare no competing interests.

\appendix

\section{Quasi-Newtonian Cosmological Formulation}  \label{App:QN_Cosm} 
Quasi-Newtonian (QN) cosmology is premised on the depiction of the Universe as a spherically symmetric distribution of points (galaxies or clusters). A given test point at a physical radius $r(t) = a(t) \ell_0$, with $\ell_0$ a fixed fiducial scale, follows the Newtonian energy conservation relation, which for the potential $V(r)$ in Eq.\,(\ref{cond0}) is 
\be \label{QN-consv}
\fr 1 2 \,{\dot r}^2 - \fr{4 \pi G} 3 \,\rho \, r^2 F(r) = E \,,
\ee
where $E$ denotes the total energy per unit mass, and $\rho = 3M/(4 \pi G r^3)$ is the total mass-energy density of the distribution, that sources $V(r)$. 

It is often convenient to express $E \equiv - k c^2/2$, where $k$ is a dimensionless constant. Then, with $F(r) = F_3(r)$, denoting $r_i = a_i \ell_0$ and $r_j' = a_j' \ell_0$ in Eq.\,(\ref{F3}), a rearrangement of terms reduces Eq.(\ref{QN-consv}) to 
\be \label{MFE0}
H^2 = \fr{8 \pi G \rho}{3 c^2} \prod_{i=1}^n \!\left(1 - \fr{a_i} a\right)^{\nu_i} \prod_{j=1}^{n'} \!\left(1 - \fr a {a_j'}\right)^{\nu_j'} \!- \frac{kc^2}{a^2}\,.
\ee
Evidently, the zeroes in $H$ are crossed at $a_i$ and $a_j'$ as the physical radius $r = a \ell_0$ sweeps through them. The cosmic evolution is {\em cyclic}, with each cycle commencing at a particular $a_i$ and ending at the corresponding $a_j'$. They are therefore the (early) {\em bounce} and the (future) {\em turnaround} points in a given cycle.

Now, a perfect fluid with energy density $\rho$ and in general a non-zero pressure $p$ can be accommodated in the QN formalism by demanding the adiabatic first law $dU + p\, d\cV = 0$ to hold, where $\cV = 4\pi \ell_0^3 a^3/3$ is the spherical volume and $U = \rho \cV$ is the fluid's internal energy. Such a demand in fact distinguishes the QN cosmology from the purely Newtonian one
\cite{DS-VarG}.
For each component in a collection of such fluids (say $N$ in number and presumably non-interacting) with individual energy densities $\rho_{_I}$ and EoS parameters $\sw_{_I}$, one consequently obtains the continuity equation 
$\dot\rho_{_I} + 3H (1 + \sw_{_I}) \rho_{_I} = 0$. 
This integrates to give 
$\rho_{_I} = \rho_{_{I0}} a^{-3 (1 + \sw_{_I})}$, with values $\rho_{_{I0}}$ at a reference epoch, usually considered the present epoch $t = t_0$ at which $a = 1$. 
Therefore, with 
\be \label{fluid_density}
\rho = \sum_{I = 1}^N \rho_{_I} = \sum_{I = 1}^N \rho_{_{I0}} a^{-3 (1 + \sw_{_I})} \,, 
\ee 
and the stipulation $n = n' = \n_1 = n_1' = 1$, Eq.\,(\ref{MFE0}) reduces to the MFE given by the above Eq.\,(\ref{MFE1}).

Note, further, the following:
\ben
\item The fiducial scale $\ell_0$ does not appear explicitly in the cosmological equation\,(\ref{MFE0}) --- it only calibrates the map $r_i(t) \to a_i(t)$ and as such gets absorbed into the parameters $a_1$ and $a_1'$.
\item The dimensionless constant $k = - 2E/c^2$ vanishes for $E = 0$ and can be rescaled to $\pm 1$ for $E \lessgtr 0$, irrespective of the stipulation of the cosmic scale factor $a = 1$ at the present epoch $t = t_0$
\cite{DS-VarG}.
As such, from a purely mathematical standpoint, $k$ is the analogue of the spatial curvature constant in standard (relativistic) cosmology.
\een
%

\section{Likelihood Analysis Methodology}
\label{App:Lik_Method}
For a given dataset of size $N_{\rm dat}$, and a parameter vector $\boldsymbol{\th}$ characterising a model $\cM$, the likelihood function is in general given by
$\cL(\boldsymbol{\th}) \propto \exp\left(- \chi^2(\boldsymbol{\th})/2\right)\,$, where 
\be \label{chi2def}
\chi^2 := \sum_{i,j=1}^{N_{\rm dat}} \D V_i \, C^{-1}_{ij} \, \D V_j,
\ee
with  
$\D V_i = V_{\rm obs}(z_i) - V_{\rm th}(z_i,\boldsymbol{\th})$,
the difference between the observed and theoretical values of the cosmological observable at a particular redshift $z_i$, and $C_{ij}$ the associated covariance matrix. The maximization of $\cL$ evidently implies the minimization of $\chi^2$ with respect to the parameters in the vector $\cL(\boldsymbol{\th})$ (say, of dimension $N_{\rm par}$).  

The goodness of fit, for the model $\cM$, is quantified by
$\chi^2_{\rm dof} \equiv \chi^2_{\rm min}/N_{\rm dof} \,$,
where $\chi^2_{\rm min}$ is the minimized value of $\chi^2$ and $N_{\rm dof} = N_{\rm dat} - N_{\rm par}$ is the number of degrees of freedom. 
%
The model comparison is generally carried out by resorting to the standard Akaike and Bayesian Information Criteria, defined respectively by 
$\, {\rm AIC} = \chi^2_{\rm min} + 2 N_{\rm par} N_{\rm dat}/(N_{\rm dof} - 1)\,$
and
$\, {\rm BIC} = \chi^2_{\rm min} + N_{\rm par} \ln N_{\rm dat}\,$,
%
%
as well as taking into account the Bayes factor $\ln \cZ$ where $\cZ = P(N_{\rm dat}|\cM)$ is the Bayesian evidence for the model $\cM$ sampled with $N_{\rm dat}$ observational datapoints.

For the statistically independent datasets we consider in this paper, the total $\chi^2$ is given by 
\be \label{chi2tot}
\chi^2_{\rm tot} = \chi^2_{\rm CC} + \chi^2_{\rm SN} + \chi^2_{\rm BAO} + \chi^2_{\rm CMB} + \chi^2_{\rm BBN} \,,
\ee 
where $\chi^2_{\rm CC}$, $\chi^2_{\rm SN}$ and $\chi^2_{\rm BAO}$ are the low-$z$ observational contributions from the Cosmic Chronometers (CC), the Pantheon$^+$ SNIa and the BAO, respectively, while $\chi^2_{\rm CMB}$ and $\chi^2_{\rm BBN}$ are the contributions from the high-$z$, CMB and BBN, observations respectively. The details of obtaining these contributions are as follows:

%
%
%
%
%
%
%
%
%
%
%

%
\bit 
\item $\chi^2_{\rm CC}$ is computed assuming diagonal uncertainties in the Monjo (2024) compilation of $51$ measurements of $H(z)$ using the CC
\cite{Monjo2024}.
\item $\chi^2_{\rm SN}$ is computed from the $1701$ datapoints of the Pantheon$^+$ SNIa sample using the corresponding statistical plus systematic covariance matrix 
\cite{Brout2022},
with the absolute magnitude offset $\D M$ treated as a nuisance parameter and sampled with a flat prior. 
\item $\chi^2_{\rm BAO}$ is computed from a total of $15$ measured distance ratios $D_H/r_d$, $D_M/r_d$ and $D_V/r_d$, from DESI DR1, SDSS MGS, SDSS DR7, and eBOSS 
\cite{Eisen2005,eBOSS2020,DESI2024},
where 
\bea 
&& D_H(z) = \fr c {H(z)}\,, ~~ D_M(z) = c \int_0^z \fr{dz'}{H(z')}\,, ~ \text{and} \nn\\ && D_V(z) = \left[z D_H(z) D_M^2(z)\right]^{1/3}
\eea  
are the radial Hubble distance, transverse comoving distance and volume-averaged distance, respectively, and $r_d$ is the sound horizon at the baryon drag epoch, which is treated as a free parameter throughout the analysis.
\item $\chi^2_{\rm CMB}$ is computed by considering the compressed Planck 2018 distance-prior likelihood, based on the mean values of the quantities $(100\,\o_b,\,100\,\th_\ast,\,\cR,\,\o_c)$ and the corresponding correlation matrix, following ref.
\cite{Arendse2020},
where $\o_b$ and $\o_c$ respectively denote 
the physical baryon density and the physical CDM density at the present epoch, $\th_\ast = r_s(z_\ast)/D_M(z_\ast)$ is the angular size of the sound horizon at the photon-decoupling redshift $z_\ast$, and $\cR  = \sq{\O_{m0}} H_0 D_M(z_\ast)/c$ is the CMB shift parameter characterizing the comoving distance to the surface of the last scattering. 
%
The photon-decoupling redshift $z_\ast$ is computed using the Hu-Sugiyama fitting formula
\cite{HuSg1996},
and consequently $r_s(z_\ast)$, the comoving sound horizon scale, is determined by numerically integrating the background expansion rate, which includes contribution from the radiation component consisting of photons and relativistic neutrinos with the standard effective number of species $N_{\rm eff} = 3.046$. The physical baryon density $\o_b$ is sampled as an independent parameter, alongside the other parameters.
\item $\chi^2_{\rm BBN}$ is computed from the primordial helium-4 mass fraction, $Y_p$, and the deuterium-to-hydrogen number ratio, $D/H$, obtained from the PRIMAT-2021 Monte Carlo fitting tables, as functions of 
$\o_b$ and the effective change in the number of relativistic species, $\D N_{\rm eff}$
\cite{Pitrou2021}.
The modified expansion history is mapped onto $\D N_{\rm eff}$ at the neutron-proton freeze-out, whence the corresponding theoretical predictions for $Y_p$ and $D/H$ are obtained and compared with the observed values
\cite{Aver2015,Cooke2018}.
%
%
\eit
%


\end{document}